\documentclass[aps,pre,twocolumn,notitlepage,superscriptaddress,nogroupedaddress]{revtex4-2}

\usepackage{xcolor,graphicx}
\usepackage{amssymb,amsmath,graphicx}
\usepackage{mathtools}
\usepackage[utf8]{inputenc}
\usepackage{comment}
\usepackage{braket}
\usepackage[normalem]{ulem}
\usepackage{hyperref,url}

\usepackage{cleveref}
\usepackage{physics}

\begin{document}

\title{Generalized Landauer bound from absolute irreversibility}

\author{Lorenzo Buffoni}
\affiliation{Department of Physics and Astronomy, University of Florence, 50019 Sesto Fiorentino, Italy}

\author{Francesco Coghi}
\affiliation{Nordita, KTH Royal Institute of Technology and Stockholm University, Hannes Alfvéns v\"{a}g 12, SE-106 91 Stockholm, Sweden}

\author{Stefano Gherardini}
\affiliation{Istituto Nazionale di Ottica -- CNR, Area Science Park, Basovizza, I-34149 Trieste, Italy}
\affiliation{SISSA, via Bonomea 265, 34136 Trieste, Italy}
\affiliation{LENS, University of Florence, via Carrara 1, I-50019 Sesto Fiorentino, Italy}

\begin{abstract}
In this work, we introduce a generalization of the Landauer bound for erasure processes that stems from absolutely irreversible dynamics. Assuming that the erasure process is carried out in an absolutely irreversible way so that the probability of observing some trajectories is zero in the forward process but finite in the reverse process, we derive a generalized form of the bound for the average erasure work, which is valid also for imperfect erasure and asymmetric bits. The generalized bound obtained is tighter or, at worst, as tight as existing ones. Our theoretical predictions are supported by numerical experiments and the comparison with data from previous works.
\end{abstract}

\maketitle


\section{Introduction}

The Landauer principle~\cite{landauer1961irreversibility} is a cornerstone of the thermodynamics of computation. It establishes a bound to the average minimum energy cost $\langle \mathbb{W} \rangle$ needed to erase a bit of information at temperature $T$. Specifically, the Landauer bound reads as
\begin{equation}
  \langle \mathbb{W} \rangle \geq k_B T \log 2 
\end{equation}
where $k_B$ is the Boltzmann constant.
From its introduction, the Landauer principle and its generalizations have been extensively studied both theoretically~\cite{DasPRE2014,gavrilov2017erasure,ProesmansPRL2020,van2022finite} and experimentally~\cite{BerutNature2012,BerutEurophysics,JunPRL2014,HoangPRL2018,ciampini2021experimental,DagoPRL2021}. In particular, the emergence of the bound from microscopic dynamics \cite{manzano2023thermodynamics}, its applications to non-ideal erasure protocols~\cite{ma2022minimal} and their limits~\cite{lee2022speed,buffoni2022third,buffoni2023cooperative,scandi2022minimally} are topics of ongoing research.

From earlier works~\cite{MurashitaPREnonequilibrium,buffoni2022spontaneous,coghi2022convergence}, we know that a dynamical system under certain conditions may exhibit \emph{absolute irreversibility}. Formally, this means that there are trajectories of the forward dynamics that can never occur while the corresponding time-reversal trajectories are realized with a non-zero probability. This condition can be formulated also from an operational point of view by postulating the existence of a regime of parameters (corresponding, e.g., to low temperature or high-transition energies) such that some particular trajectories of a process $\mathbf{Z}_t$ evolving from time $0$ to $\tau$ (thus, $t\in[0,\tau]$) are suppressed in the transient. In other words, the probability of observing a certain forward-in-time trajectory $\omega_{\rightarrow}$, given a certain initial distribution for the process, is zero, whereas the probability of observing the corresponding time-reversal trajectory $\omega_{\leftarrow}$ is greater than 0.
Arguably, the condition of suppressing some transitions between states is in many cases important for information processing~\cite{manzano2023thermodynamics,wolpert2020thermodynamics}. As an example, in building a bona fide bit of memory, the transitions between the macrostates representing the logical states $0$ and $1$ shall not be allowed once the system is in the stationary state. Otherwise, the information encoded in the bit would not be preserved. Hence, in real implementations, it is plausible that highly accurate erasure processes can be realized with a protocol that works out-of-equilibrium in a regime of absolute irreversibility.

In this paper, we provide a new perspective in deriving and interpreting the Landauer bound for erasure processes, by including absolute irreversibility from first principles. Specifically, we are going to obtain a generalization of the bound valid beyond paradigmatic cases, such as the one of a perfectly-erased symmetric bit. 
We show that, if the erasure process is carried out in an absolutely irreversible way, the bound for the average erasure work is tighter than the one experimentally achieved in~\cite{BerutNature2012,BerutEurophysics} in case of imperfect erasure, and as tight as the bound in~\cite{gavrilov2017erasure} for the erasure of a bit in an asymmetric double-well potential. Eventually, we test the generalized Landauer bound through numerical simulations of a ferromagnetic Ising model.

\section{Generalized Landauer bound}
\label{sec:GLB}

We here introduce a novel formulation of the Landauer bound that stems from absolute irreversibility~\cite{MurashitaPREnonequilibrium}. We assume that (i) the initial density of states of the forward process $\mathbf{Z}_t$ is a thermal distribution with inverse temperature $\beta = (k_{B}T)^{-1}$, (ii) the initial density of the time-reversal process is also a thermal distribution with the same inverse temperature of the forward process, and that (iii) the final state of the process asymptotically equilibrates with the environment at inverse temperature $\beta$. Noticeably, no assumptions are set on the Hamiltonian of the system whose time-dependence, in principle, can be arbitrary.

The fluctuating entropy production $\mathbb{S}$~\cite{LebowitzJSP1999,Barato2015} associated to a trajectory of the process $\mathbf{Z}_t$, linking initial and asymptotic thermal states, is given by
\begin{equation}\label{eq:equilibrium_entropy}
    \mathbb{S}(\omega_{\rightarrow}) = \beta (\mathbb{W}(\omega_{\rightarrow})-\Delta F)
\end{equation}
where $\mathbb{W}$ is the stochastic work performed on/by the system and $\Delta F$ is the free-energy difference between initial and final state~\cite{seifert2012stochastic}.

Classically, under the validity of assumptions (i)-(iii), we know that the average $\langle e^{-\beta(\mathbb{W} - \Delta F)}\rangle=1$. This is the well-known Jarzynski equality~\cite{Jarzynski97}, which holds normally but may break in the presence of absolute irreversibility. This is showed in the seminal paper~\cite{MurashitaPREnonequilibrium} where the average over all trajectories of the system for an absolute irreversible dynamics reduces to
\begin{equation}\label{eq:average_entropy_Ueda}
    \langle e^{-\beta(\mathbb{W}-\Delta F) } \rangle = 1 - \lambda 
\end{equation}
where the constant factor $\lambda \in [0,1)$ is the fraction of absolute irreversible trajectories, i.e., trajectories that have zero occurrence probability in the forward process but non-zero in the reversed dynamics. As a remark, we have to stress that, for a given Hamiltonian, absolute irreversibility is not always valid for any initial state of the forward dynamics and/or any dynamical process. The paradigmatic example we are going to consider in the following is the erasure of logical bits.

Let us thus consider erasure protocols encoded in a cyclic process, i.e. processes driven in a time $\tau$ by a time-dependent Hamiltonian $H(t)$ with $H(0)=H(\tau)$, and carried out at a constant temperature $\beta$. Using the microscopic definition of the free-energy $F = - k_{B}T \log Z$, with $Z$ denoting the partition function that depends only on the temperature and the Hamiltonian of the system, 
%
%
one can determine that for any cyclic process $\Delta F = 0$. In this way, the application of the Jensen's inequality to (\ref{eq:average_entropy_Ueda}) leads to the following lower bound for the average erasure work: 
\begin{equation}\label{eq:generalized_Landauer}
    \langle \mathbb{W} \rangle \geq - k_B T \log(1-\lambda) \,.
\end{equation}
In case of erasure protocols, the right-hand-side (r.h.s.) of the inequality (\ref{eq:generalized_Landauer}) generalizes the standard Landauer bound $\langle \mathbb{W} \rangle \geq k_{B}T \log(2)$ including absolute irreversibility. We also remark that if the logical states, classically corresponding to macrostates, are initially sampled from a non-thermal distribution, then also the free-energy difference $\Delta F$ has to be included in the r.h.s.~of the inequality (\ref{eq:generalized_Landauer}) as $\langle \mathbb{W} \rangle \geq \Delta F - k_{B}T \log(1-\lambda)$.

\subsection{Erasure protocol with $2$ logical states}

We specialize our analysis to an erasure protocol with two logical states $0$ and $1$, setting the reset state to $1$ (were the reset state be $0$, the argument would be the same, simply mirrored). From a macroscopic point of view, only $4$ possible transitions between the logical states are allowed. By marking $k$ the initial logical state and $\ell$ the final one, with $k,\ell\in\{0,1\}$, we define the normalized joint probability $P_{k,\ell}$ to be the probability of initializing the protocol at $k$ and ending it at $\ell$. Thanks to conditioning, we can write $P_{k,\ell}=P_{k}P_{\ell|k}$, with $P_{k}$ the probability to initialize the bit in the logical state $k$ and $P_{\ell|k}$ the conditional probability of reaching the logical state $\ell$ from initialization at $k$. Such an erasure protocol is customarily associated with the following joint probabilities:
\begin{equation}\label{eq:p_ij_Ising}
    \left\{ \begin{array}{lr}
    P_{1,0} = 0 \\
    P_{1,1} = \eta \\
    P_{0,0} = (1-q_s)(1-\eta) \\
    P_{0,1} = q_s(1-\eta)  \\
    \end{array}\right. 
\end{equation}
where we impose $P_{0|1}=0$, such that $P_{1|1}=1-P_{0|1}=1$, $P_1 = \eta$ denotes the probability to initialize the bit in the erased logical state $1$, and $q_s$ is the success probability of correctly erasing the bit through the transition $0 \rightarrow 1$. It is worth observing that the joint probability $P_{k,\ell}$ in \eqref{eq:p_ij_Ising} reproduces already known scenarios of erasure protocols, such as the \emph{perfect erasure} of \emph{symmetric bits} that is indeed obtained by setting $\eta=0.5$ and $q_s=1$: $P_{0,0}=P_{1,0}=0$ and $P_{1,1}=P_{0,1}=0.5$. Moreover, at this level of description, no assumptions on the microscopical details of the erasure protocol and its dynamics are made.

The condition of taking $P_{0|1}=0$ should be a requisite of any erasure protocol; in fact, once the bit is in the erasure state, one would like the occurrence of no further transitions. Incidentally, imposing that a conditional probability is equal to $0$ is tantamount of introducing absolute irreversibility in this specific protocol. Notice that this should not be intended as a peculiar feature of $P_{1,0}$, but in principle of every suppressed transition.

For erasure protocols with $2$ logical states, we can then rewrite (\ref{eq:average_entropy_Ueda}) as
\begin{equation}\label{eq:jarz-absolute}
    \langle e^{-\beta\mathbb{W} } \rangle = 1 - q_s(1-\eta) \, .
\end{equation}
The lower bound for the average erasure work then reads as
\begin{equation}\label{eq:Landauer_Ising}
    \langle \mathbb{W} \rangle \geq - k_{B}T \log\left( 1 - q_s(1-\eta) \right) \, .
\end{equation}
This result comes from reversing the dynamics in \eqref{eq:p_ij_Ising} by driving the system backward using the time-reversed Hamiltonian $H(\tau-t)$ as in \cite{Jarzynski2006}. This entails that
\begin{equation}\label{eq:p_ij_Ising_reverse}
    \left\{ \begin{array}{lr}
    \Tilde{P}_{1,0} = q_s(1-\eta) \\
    \Tilde{P}_{1,1} = \eta \\
    \Tilde{P}_{0,0} = (1-q_s)(1-\eta) \\
    \Tilde{P}_{0,1} = 0 \\
    \end{array}\right. \, , 
\end{equation}
where $\Tilde{P}$ denotes the probabilities of the transitions in the backward process. We thus compute the fraction $\lambda$ of absolute irreversible trajectories that, by definition, corresponds to the probability of a transition occurring in the backward process that has zero probability of occurrence in the forward process. Hence, under the assumptions above, we get $\lambda = q_s(1-\eta)$ by comparing \eqref{eq:p_ij_Ising} and \eqref{eq:p_ij_Ising_reverse}. In this way, inserting $\lambda = q_s(1-\eta)$ in \eqref{eq:generalized_Landauer} leads to \eqref{eq:Landauer_Ising}. It is worth noting that the fraction $\lambda$ of absolute irreversible trajectories generated by the microscopic system dynamics are implicitly inferred from the macroscopic transition probabilities $P$ and $\Tilde{P}$ between macrostates, here represented by the logical states.

In the case of a perfect erasure of symmetric bits ($\eta=0.5$ and $q_s=1$), \eqref{eq:Landauer_Ising} becomes the standard Landauer bound $\langle \mathbb{W} \rangle \geq k_{B}T \log(2)$. In a similar fashion, by considering $q_s=0$, viz.\ leaving the system unperturbed from the initial time until the end of the protocol ($P_{0,1}=P_{1,0}=0$, $P_{1,1}=\eta$, $P_{0,0}=1-\eta$), we correctly get $\langle \mathbb{W} \rangle \geq 0$, in agreement with the second law of thermodynamics. 

\section{Simulations of a ferromagnetic Ising model}
\label{sec:Ising_ferromagnet}

\begin{figure*}
    \centering
    \includegraphics[width=\linewidth]{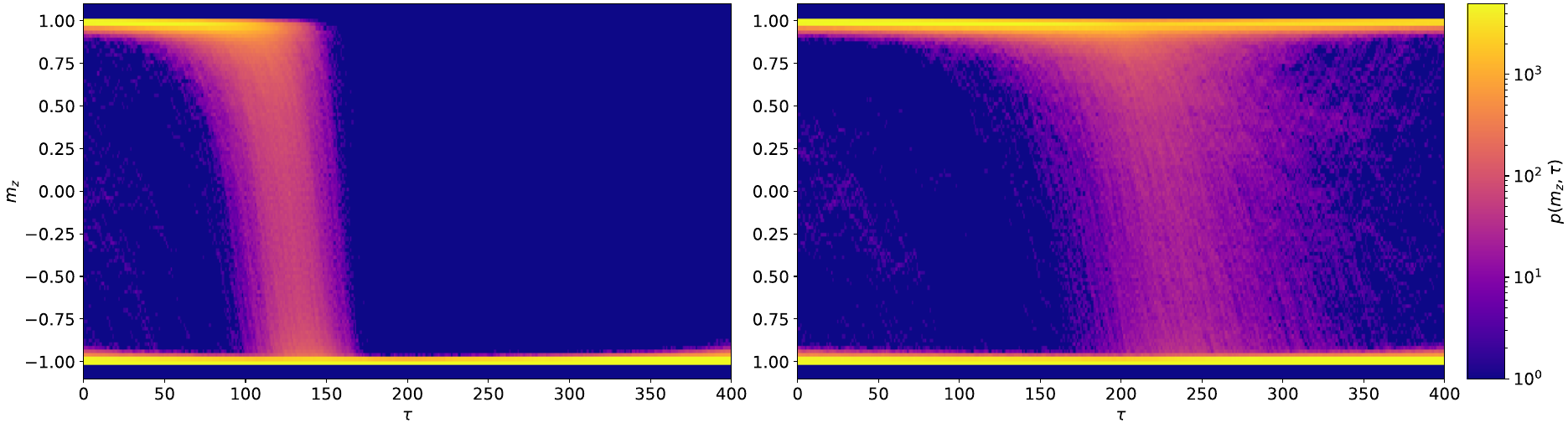}
    \caption{
    Density plot of 10000 normalized magnetization trajectories for an erasure protocol applied to a macroscopic bit realized by a $16\times 16$ Ising ferromagnet. In the left panel, a high local field $h_{max}=1$ is used to erase the initial bit, such that $q_s=1$. In the right panel, a smaller field $h_{\rm max}=0.43$ is employed, thus resulting in an erasure with $q_s=0.45$. In both cases, the initial bits, corresponding to all spins up or down of the Ising ferromagnet, are distributed such that $\eta \approx 0.5$ with a parameter $\beta=1.5$. The duration of the protocol, following a Glauber dynamics, is fixed at $\tau=400$ time steps.}
    \label{fig:ising-sim}
\end{figure*}

In this section we test the new lower bound (\ref{eq:generalized_Landauer}) for the average erasure work on a $16\times 16$ Ising ferromagnet below the critical temperature $T_c$. The Hamiltonian of the model is
\begin{equation}\label{eq:Ising_Hamiltonian}
H(t) =  h(t) \sum_i \sigma_i^z - J \sum_{[ i,j ]} \sigma_i^z \sigma_j^z 
\end{equation}
where $\sigma_{k}^{z}$ denotes the spin value along the $z$-axis at the site $k$, $J$ is the coupling constant, and $[ i,j ]$ indicates all pairs $(i,j)$ of sites with nearest-neighbour interactions. The critical temperature of the ferromagnet behaving with the isotropic 2D Ising Hamiltonian (\ref{eq:Ising_Hamiltonian}) is $T_c = 2J/( k_B \log(1+\sqrt{2}) )$  \cite{baxter2016exactly}. The Ising ferromagnet represents a \emph{macroscopic bit} by encoding the logical values $0$ and $1$ in the positive and negative magnetization states (respectively, all spins down or up). The erasure of the bits is enabled by the local field $h(t)$ that, from $h_0=0$ at the beginning of the protocol, is linearly increased up to a certain value $h_{\rm max}$. Then, it is brought back to zero in a time $\tau$, in order to make the protocol cyclical (such that $\Delta F = 0$) and the value of the resulting bit, either $0$ or $1$, stationary. Here, it is worth observing that considering the protocol as  cyclic is a fundamental property 
%
%
to make the bound applicable in practical physical contexts. In particular, at the end of the erasure protocol, the logical system has to be still encoded in a bit with the same properties of the one employed before the application of the protocol. In other terms, we need that the macrostates associated to the logical system are the same for any bit erasure. This requirement is not generally possible in the case a non-cyclical protocol is applied.
%

As we can see from Fig.~\ref{fig:ising-sim}, depending on the value of $h_{\rm max}$, one can perform erasure with different probabilities $q_s\leq 1$. Each initial logical bit $0$ and $1$ is evolved according to Glauber dynamics~\cite{glauber1963time}, with inverse temperature $\beta=1.5$ and for a duration of $\tau=400$ time steps, so as to produce trajectories $\omega_{\rightarrow} = \{\sigma_1(t),\sigma_2(t),\cdots,\sigma_L(t)\}_{t\in[0,\tau]}$ of the spin dynamics with $L = 256$. The erasure protocol that is realized with this spin dynamics is repeated $N = 10^6$ times. For each resulting trajectory $\omega_{\rightarrow}$, we firstly compute the global magnetization 
\begin{equation}
    M(\omega_{\rightarrow}) = \frac{1}{L}\sum_{i=1}^{L}\sigma_i(t)
\end{equation}
at any time $t$, with $M \in [0,1]$. Then, we derive the stochastic work $\mathbb{W}(\omega_{\rightarrow})$ in the time interval $[0,\tau]$, which is obtained from $M(\omega_{\rightarrow})$ through the following relation:
\begin{equation}
    \mathbb{W}(\omega_{\rightarrow}) = \int_{0}^{\tau} dt \, \dot{{h}}(t)  {M}(\omega_{\rightarrow}) \, .
\end{equation}
The work statistics is then used to compute the empirical mean $\langle e^{-\beta \mathbb{W}} \rangle_N$ over a sample of $N$ trajectories of $\omega_{\rightarrow}$. The number $N$ of trajectories is taken sufficiently large such that the empirical mean $\langle\cdot\rangle_N$ well approximates the mathematical average $\langle\cdot\rangle$.

Absolute irreversibility can arise for a given set of parameters values that define the dynamics of the implemented protocol. In our case-study, the relevant parameters are $h_{\rm max}$, $q_s$, $\beta$ and $\tau$. Also notice that, in the following, we aim at numerically verifying the equality (\ref{eq:jarz-absolute}) and not the inequality (\ref{eq:Landauer_Ising}) for two main reasons. Firstly, checking inequalities is harder since one has to find a regime of parameters in which the dynamics potentially saturates the bound and this may not even be possible. Secondly, the lower bound in (\ref{eq:Landauer_Ising}) directly derives from the application of the Jensen's inequality on (\ref{eq:jarz-absolute}); thus, verifying (\ref{eq:jarz-absolute}) implies verifying the bound.

We show in Table \ref{tab:simple_table_1} the results obtained from the numerical simulations by setting $\beta=1.5$, $\tau=400$ and with varying the field $h_{\rm max}$ such that the protocol works in the imperfect erasure regime. Specifically, by lowering the value of $h_{\rm max}$, the erasure accuracy $q_s$ decreases quite noticeably. Since we have access to the trajectories $\omega_{\rightarrow}$ of the forward dynamics, we can compute the average exponentiated work and thus check if (\ref{eq:average_entropy_Ueda}) is verified with $\lambda = q_s(1-\eta)$, as predicted by our theory. In all tested cases reported in Table \ref{tab:simple_table_1}, the values of $\langle e^{-\beta \mathbb{W}} \rangle_N$ and $\lambda$ are quite close, thus substantiating the validity of the lower bound in (\ref{eq:Landauer_Ising}).

\begin{table}[htbp]
    \centering
    \begin{tabular}{|c|c|c|c|c|}
        \hline
        $h_{max}$ & $q_s$ & $\eta$ & $\langle e^{-\beta \mathbb{W}} \rangle_N$ & $1-q_s(1-\eta)$ \\
        \hline
        1 & 1.0 & 0.497 & 0.502 (0.006) & 0.497 \\
        \hline
        0.5 & 0.8 & 0.499 & 0.617 (0.022) & 0.599 \\
        \hline
        0.4 & 0.301 & 0.501 & 0.903 (0.053) & 0.849 \\
        \hline
    \end{tabular}
    \caption{Results of numerical erasure experiments with different values of $h_{max}$, starting from a symmetric distribution $\eta \approx 0.5$. The validity of the equality in (\ref{eq:jarz-absolute}) for the average erasure work (with standard deviation reported in parentheses) is tested by matching the values in the last two columns of the table.}
    \label{tab:simple_table_1}
\end{table}

It is worth noting that the trajectories of the erasure protocols in Table \ref{tab:simple_table_1} are initially sampled with $\eta \approx 0.5$ due to the symmetry of the Ising ferromagnet configurations at the equilibrium.
However, we can check numerically the validity of the equality in (\ref{eq:jarz-absolute}) for different values of $\eta$, by post-selecting the trajectories of the dynamics. In particular, one can take only the trajectories realizing a successful erasure. For the post-selected trajectories $\hat{\omega}_{\rightarrow}$, we define the success rate $\hat{q}_s$ that is equal to $1$ by definition, and the initial distribution of the sampled trajectories is thus shifted becoming $\hat{\eta}$. As a result, one implicitly gets $\langle e^{-\beta\hat{\mathbb{W}}} \rangle_N = \hat{\eta}$. Our prediction for the case of asymmetric bit erasure finds confirmation in Table \ref{tab:simple_table_2}.

\begin{table}[htbp]
    \centering
    \begin{tabular}{|c|c|c|c|}
        \hline
        $h_{max}$ & $\hat{q}_s$ & $\hat{\eta}$ & $\langle e^{-\beta\hat{\mathbb{W}}} \rangle_N$  \\
        \hline
        1 & 1.0 & 0.497 & 0.502 (0.06) \\
        \hline
        0.5 & 1.0 & 0.555 & 0.556 (0.01) \\
        \hline
        0.4 & 1.0 & 0.766 & 0.768 (0.03) \\
        \hline
    \end{tabular}
    \caption{Test of (\ref{eq:jarz-absolute}) for the erasure of asymmetric bits. The results are obtained by post-selecting the trajectories of the simulated erasure protocol to single-out only the ones that successfully bring the Ising ferromagnet into the reset state.
    }
    \label{tab:simple_table_2}
\end{table}

\section{Tighter bound from absolute irreversibility}

We now show that the lower bound (\ref{eq:generalized_Landauer}) for the average erasure work is tighter, or as tight as other bounds existing in the literature. Specifically, we are going to compare (\ref{eq:generalized_Landauer}) with the lower bound derived in ~\cite{BerutNature2012,BerutEurophysics} for the \emph{imperfect erasure}, and with the lower bound in \cite{gavrilov2017erasure} for the erasure of a bit in an \emph{asymmetric} double-well potential. All these results are in agreement with the second law of thermodynamics, and they show that the average work needed to erase a bit can reach the Landauer limit $k_B T\log(2)$ when symmetric bits, i.e., bits where the two logical states are equiprobable at equilibrium, are taken into account. We further notice that, due to systematic errors, the lower bound in \cite{BerutNature2012,BerutEurophysics} can be smaller or larger than the Landauer limit if the probability $p_s$ to successfully erase the bit is less than $1$. For similar reasons, also in \cite{gavrilov2017erasure} it is proved that in case of asymmetry the average work for erasure can go below than the Landauer limit.

\begin{figure}
    \centering
    \includegraphics[width=0.99\linewidth]{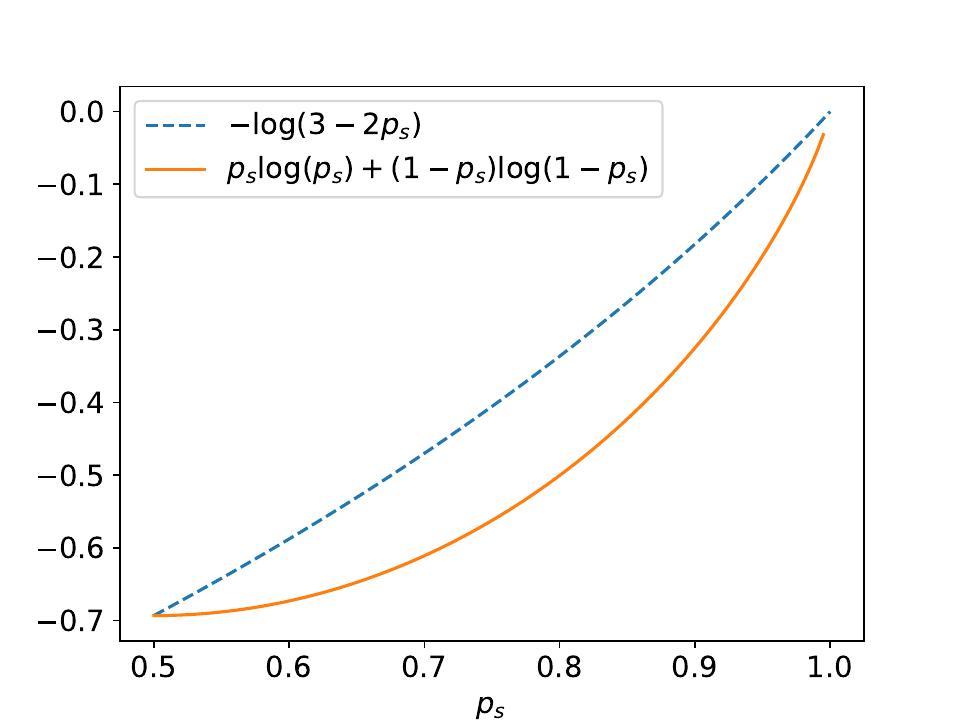}
    \caption{We plot the correction to the Landauer bound for the case of imperfect erasure with probability $p_s$ using both our approach stemming from absolute irreversibility (blue dashed line) and the one in \cite{BerutNature2012,BerutEurophysics} (orange solid line). 
    }
    \label{fig:bound_compar}
\end{figure}

We consider erasure protocols with $2$ logical states. In \eqref{eq:Landauer_Ising} we add and substract $\log 2$ to get
\begin{equation}\label{eq:our_bound_kTlog2}
    \langle \mathbb{W} \rangle \geq k_{B}T \left[\log(2) - \log\left( 2 - 2q_s(1-\eta) \right)\right]\,.
\end{equation}
We now compare the r.h.s.~of \eqref{eq:our_bound_kTlog2} with the lower bound in \cite{BerutNature2012,BerutEurophysics}, which reads
\begin{equation}\label{eq:work_bound_Ciliberto}
    \langle \mathbb{W} \rangle \geq k_B T[\log(2) + p_s\log(p_s) + (1-p_s)\log(1-p_s)] 
\end{equation}
where the probability $p_s$ is operationally provided by the fraction of trajectories that end in the chosen reset state (the logical state $1$ in this paper). Notice that $\log(2) + p_s\log(p_s) + (1-p_s)\log(1-p_s)$ quickly decreases to $0$ when $p_s$ is smaller than $1$; we show in the following that, by using the notion of absolute irreversibility, a tighter lower bound for the average erasure work can be obtained. Indeed, in our case-study we have $p_s = \sum_{k}P_{k,1} = \eta + q_s(1-\eta) < 1$. This entails that, for a symmetric bit ($\eta=0.5$), $q_s(1-\eta)=p_s-\frac{1}{2}$, with the result that the inequality (\ref{eq:our_bound_kTlog2}) stemming from absolute irreversibility reads as
\begin{equation}\label{eq:work_bound_absolute_irr_ps}
\langle \mathbb{W} \rangle \geq k_{B}T \left[\log(2) - \log\left( 3 - 2p_s \right)\right]\,.
\end{equation}
The lower bound for the erasure work from absolute irreversibility is tighter than the one provided by \cite{BerutNature2012,BerutEurophysics} if the r.h.s.~of \eqref{eq:work_bound_absolute_irr_ps} is larger than the r.h.s.~of \eqref{eq:work_bound_Ciliberto}, i.e., if 
\begin{equation}\label{eq:comparison_1}
     - \log\left( 3 - 2p_s \right) \geq  p_s \log(p_s) + (1-p_s)\log(1-p_s)
\end{equation}
for any $p_s$ in the interval of interest $[0.5,1]$. Remarkably, the inequality (\ref{eq:comparison_1}) is always satisfied, as clearly showed in Fig.~\ref{fig:bound_compar}.
Furthermore, this finding confirms the experimental observation that the average work spent in realizing an imperfect erasure in \cite{BerutNature2012} is significantly higher than the estimate given by \eqref{eq:work_bound_Ciliberto}, albeit lower than $k_B T\log (2)$.

As mentioned before, the lower bound in \cite{gavrilov2017erasure} well describes (theoretically and experimentally) the average work needed to erase a bit whose logical values are encoded in asymmetric equilibrium states. Notably, the result in \cite{gavrilov2017erasure} is exactly recovered from the lower bound of \eqref{eq:our_bound_kTlog2} by setting $q_s=1$ (perfect erasure). In such a case, indeed, the lower bound of the average erasure work in \cite{gavrilov2017erasure} becomes
\begin{equation}\label{eq:bound_Gavrilov}
    \langle \mathbb{W} \rangle \geq k_{B}T \left[\log(2) - \log\left( 2\eta \right)\right] 
\end{equation}
where $\eta$ is the asymmetry in the initial distribution of the bits. In \cite{gavrilov2017erasure}, $\eta$ is realized in an asymmetric double-well potential, and encoded in the fraction of states that are in the reset state at the equilibrium. \eqref{eq:our_bound_kTlog2} and \eqref{eq:bound_Gavrilov} clearly provide the same expression for $q_s=1$. It is worth noting that if $\eta>0.5$ the correction to $k_{B}T \log(2)$ is \emph{negative}, with the result that one can go below the Landauer bound. Instead, if $\eta<0.5$ the correction is \emph{positive}.

\section{Conclusions}

In this paper, under the hypothesis of absolute irreversibility, we have showed how to derive a generalization of the Landauer bound for erasure protocols that can be imperfect or operate on asymmetric bits. The bound we obtain generates directly from the breakdown of fluctuation theorems in the regime of absolute irreversibility. The lower bound \eqref{eq:Landauer_Ising} of the average erasure work is tighter or, at worst, as tight as other bounds existing in the literature referring to two limiting scenarios: (i) the imperfect erasure of symmetric bits, and (ii) the perfect erasure of asymmetric bits. Notably, the bound in \eqref{eq:Landauer_Ising} can be applied to all combinations of cases (i)-(ii), and extended to multi-state logic systems, going beyond the erasure of single bits.

We think it is interesting to understand the limits upon which this generalization remains valid from both a theoretical and a practical standpoint, also by extending it to the erasure of logical quantum states. In particular, if one takes an erasure protocol that can be made absolutely irreversible \cite{funo2015quantum} (as the one in our numerical example), one can investigate in which regime absolute irreversibility holds and if and how we can perform erasure outside of the regime of absolute irreversibility.

Finally, we believe it is worth testing our results on experimental platforms that have already been employed to verify the standard Landauer bound~\cite{berut2015information,DagoPRL2021}, with an additional control over the erasure probability.

\section*{Acknowledgments}

The authors acknowledge support from the MISTI Global Seed Funds MIT-FVG Collaboration Grant ``Non-Equilibrium Thermodynamics of Dissipative Quantum Systems (NETDQS)'', and the PNRR MUR project PE0000023-NQSTI. L.B. was funded from Next Generation EU, in the context of the National Recovery and Resilience Plan, M4C2 investment 1.2. Project SOE0000098-ThermoQT. This resource was financed by the Next Generation EU [DD 247 19.08.2022]. The views and opinions expressed are only those of the authors and do not necessarily reflect those of the European Union or the European Commission. Neither the European Union nor the European Commission can be held responsible for them.

\bibliographystyle{unsrt}
\bibliography{bibliography}

\end{document}